\documentclass[12pt]{article}  
\usepackage[centertags]{amsmath}
\usepackage[english]{babel}
\usepackage{amsthm}
\usepackage{mathtools}
\usepackage{fancyhdr}  
\usepackage{setspace}   
\usepackage{caption} 

\usepackage{amsfonts}
\usepackage{amssymb}
\usepackage[pdftex]{graphicx}
\usepackage{booktabs} 
\usepackage[longnamesfirst]{natbib}
\usepackage{url}
\usepackage[ruled,linesnumbered]{algorithm2e}
\usepackage{algorithmicx,algpseudocode}
\usepackage{amssymb}
\usepackage{setspace}
\usepackage{cancel}
\usepackage{pgf}
\usepackage{tikz}
\usetikzlibrary{shapes,arrows,automata,arrows.meta}

\usepackage{multirow}

\title{ \Large \textbf{Technical Report} \\ \vspace{10mm} The Bradly-Terry Regression Trunk approach for modelling preference data with small trees. \\   }

\author{Alessio Baldassarre*$^1$, Elise Dusseldorp**$^2$, \\
Antonio D'Ambrosio***$^{3}$, Mark de Rooij**$^{4}$\\
and Claudio Conversano*$^{5}$\\
\\\small *University of Cagliari\\ 
\\\small **Leiden University\\
\\\small ***University of Naples Federico II\\
\small $^1$a.baldassarre@studenti.unica.it, $^2$elise.dusseldorp@fsw.leidenuniv.nl\\
\small $^{3}$ antdambr@unina.it, $^{4}$rooijm@fsw.leidenuniv.nl, $^{5}$conversa@unica.it}

\date{July 29, 2021}

\begin{document}

\tikzstyle{decision} = [diamond, draw, 
    text width=4.5em, text badly centered, node distance=3cm, inner sep=0pt]
\tikzstyle{block} = [rectangle, draw, text width=8cm, text centered, rounded corners, minimum height=0.5cm, minimum width=3cm]
\tikzstyle{blockyes} = [rectangle, draw, text width=3cm, text centered, rounded corners, minimum height=0.5cm, minimum width=1cm]
\tikzstyle{cloud} = [draw, circle, text width=1cm, minimum height=0.5cm, minimum width=1cm]

\maketitle
\newpage


\noindent \textbf{Abstract}.
This paper introduces the Bradley-Terry Regression Trunk model, a novel probabilistic approach for the analysis of preference data expressed through paired comparison rankings. In some cases, it may be reasonable to assume that the preferences expressed by individuals depend on their characteristics.  Within the framework of tree-based partitioning, we specify a tree-based model estimating the joint effects of subject-specific covariates over and above their main effects. We combine a tree-based model and the log-linear Bradley-Terry model using the outcome of the comparisons as response variable.  The proposed model provides a solution to discover interaction effects when no a-priori hypotheses are available. It produces a small tree, called trunk, that represents a fair compromise between a simple interpretation of the interaction effects and an easy to read partition of judges based on their characteristics and the preferences they have expressed. We present an application on a real data set following two different approaches, and a simulation study to test the model's performance. Simulations showed that the quality of the model performance increases when the number of rankings and objects increases. In addition, the performance is considerably amplified when the judges’ characteristics have a high impact on their choices. 
\\


\section{Introduction}

The analysis of preference data is ubiquitous in many scientific fields, such as social sciences, economics, political sciences, computer science, psychometrics, behavioral sciences. There are several ways to analyze preferences, mainly depending on how these are collected from a set of individuals, or judges. For example, people can express their preferences with respect to a set of items (or stimuli, or objects) by assigning a numerical value to each of them according to an ordinal scale. Sometimes, instead of assigning a numeric score to each item, people can place in order the objects by forming a list in which the preferences are stated simply by looking at the order in which each object appears in the list. This list is called ordering (or order vector), and it can be transformed into a ranking (or rank vector) when, given any arbitrary order of the set of the objects, the rank of each of them is reported. \citep{marden}. \\
Sometimes objects are presented in pairs to judges, producing the so-called paired comparison rankings: this could be the natural experimental procedure when the objects to be ranked are really similar and the introduction of others may be confusing \citep{david1969}. Given a ranking of $n_o$ objects, it is always possible to determine the relative $n_o\times (n_o-1)/2$ pairwise preferences. On the other hand, a set of $n_o \times (n_o-1)/2$ paired comparisons does not always correspond to a ranking because of the phenomenon of non-transitivity of the preferences. Such non-transitivity could be avoided by ensuring that `individuals comparisons are independent or nearly' \citep[][p. 11]{david1969}.\\
In analyzing rank data, the goal is often to find one ranking that best represents all the preferences stated by the individuals. This goal, when dealing with rank vectors, is known as the consensus ranking problem, the Kemeny problem, or the rank aggregation problem \citep{dambrosio2019}. When dealing with paired comparison rankings, the goal is to determine the probability that object $i$ is preferred to object $j$ for all the possible pairs of them: the final outcome is thus a probabilistic determination of the central ranking \citep{kendallbb1940, bradley1952, mallows1957}. \\
Finding the central ranking is a very important step when rank data are analyzed \citep{cook1982, emond2002, meila2007, d2015, amodio2016, aledo2017} either as a final analysis tool, when homogeneity among people is assumed, or as a part of a more complex analysis strategy, when heterogeneity among judges is assumed. More generally, preference rankings can be analyzed with several statistical models and methodologies, both supervised and unsupervised. Among these, there are methods based on the goodness-of-fit adaptation and probabilistic methods \citep{marden, heiserdambr}. The first category includes methods such as Principal Component Analysis \citep{carroll1972}, Unfolding \citep{coombs1950, coombs, prefscal, van2007, busing2010,dambrveraheiser}, Multidimensional Scaling \citep{heiser1981multidimensional, hooley} and  Categorical Principal Component Analysis \citep{meulman2004}. These methods are intended to describe the structure of rank data. On the other hand, the probabilistic methods can assume a homogeneous or heterogeneous distribution of judges. In the first case, they focus on the ranking process assuming solid homogeneity among the judges' preferences. In the second one, the methods are aimed at modeling the population of judges assuming substantial heterogeneity in their preferences. When homogeneity is assumed, probabilistic methods are based on the so-called Thurstonian models,  distance-based and multistage models \citep{thurstone1927, bradley1952, mallows1957, luce1959}, mixtures of Bradley-Terry-Luce models, mixtures of distance-based models \citep{croon1989, murphy2003, gormley2008exploring}, and probabilistic-distance methods \citep{d2019}.
The probabilistic methods that assume heterogeneity are based on a reasonable concept: Different groups of subjects with specific characteristics may show different preference rankings \citep{strobl2011}. Such heterogeneity can be accounted for by the introduction of subject-specific covariates, from which mixtures of known sub-populations can be estimated. In most cases, the methods that consider covariates are based either on generalized linear models \citep{chapmann1982, dittrich2000, bockenholt2001, francis2002, skrondal2003, gormley2008MOE} or recursive partitioning methods (i.e., tree-based) \citep{strobl2011, lee2010, d2016recursive, plaia2019}.\\ 
In the literature, there is relatively little work in the classification community that uses the typical properties of rankings. \cite{dittrich2000} proposed a parametric model for the analysis of rank ordered preference by means of Bradley-Terry type models when categorical subject-specific covariates are observed. Their idea was to transform the (complete) rankings data into paired comparisons, and then to apply a log-linear model for a corresponding contingency table. The authors proposed a procedure for researching the interaction effects between covariates by applying a forward selection and backward elimination procedure. This approach is well suited for hypothesis-based modeling. However, when no a priori hypotheses are known, it requires the arbitrary introduction of higher-order interactions.\\
\cite{strobl2011} proposed a tree-based classifier, where the paired comparisons are treated as response variables in Bradley-Terry models. They found a way to discover interactions when no a priori hypothesis is known, suggesting a model-based recursive partitioning where splits are selected with a semi-parametric approach by looking for instability of the basic Bradley-Terry model object parameters. The final result provides the preference scales in each group of the partition that derives from the order of object-related parameters, but it does not offer information about how the subject-specific covariates affect the judges' preferences. This semi-parametric model, therefore, returns beta coefficients neither for the main effects nor for the interaction effects between the covariates.\\

To overcome the drawbacks characterizing the works of \cite{dittrich2000} and \cite{strobl2011} we propose an alternative approach that fits a generalized linear model with a Poisson distribution by combining its main effects with a parsimonious number of interaction effects. Our proposal is framed within the Simultaneous Threshold Interaction Modeling Algorithm (STIMA) proposed by \cite{dusseldorp2010combining} and \cite{CD17} that, in the case of a numerical response, is based on the Regression Trunk Approach \citep{dusseldorp2004}. Dealing with paired comparisons, our approach combines the extended log-linear Bradley-Terry model including subject-specific covariates with the regression trunk. Thus, the proposed model is named \textit{Bradley-Terry Regression Trunk (BTRT)}. It produces an estimated generalized linear model with a log link and a Poisson distribution presenting a main effects part and an interaction effects part, the latter being composed of a restricted number of higher-order interactions between covariates that are automatically detected by the STIMA algorithm. The interaction effect part can be graphically represented in a decision tree structure, called trunk, because it is usually characterized by few terminal nodes. Hence, BTRT allows observing the preference scale in each node of the trunk and to evaluate how the probability of preferring specific objects changes for different groups of individuals. The final result is a small tree that represents a compromise between the interpretability of interaction effects and the ability to summarize the available information about the judges' preferences.

The rest of the paper is organized as follows. In Section \ref{BTM}, we give an overview of the basic Bradley-Terry model and its extension with subject-specific covariates. Next, the STIMA algorithm and the regression trunk methodology are explained in Section \ref{RT}. In Section \ref{prop} we describe BTRT and show how it can efficiently be used for the task of partitioning individuals based on their preferences. A simulation study has been carried out to investigate, in particular, on the choice of a suitable pruning rule: results are reported in Section \ref{sim}. In Section \ref{RD} we present an application of BTRT on a real data set.  Conclusions and future research directions are reported in Section \ref{conc}.

\section{The Bradley-Terry model}
\label{BTM}

The model proposed by \cite{bradley1952} is the most widely used method for deriving a latent preference scale from paired comparison data when no natural measuring scale is available \citep{strobl2011}. It has been applied in psychology and several other disciplines. Recent applications include, for example, surveys on health care, education, and political choice \citep{dittrich2006} as well as psycho-physical studies on the sensory evaluation of pain, sound, and taste \citep{choisel2007} or in prioritization of balance scorecards \citep{MBPP20}.

The paired comparison method splits the ordering process into a series of evaluations carried out on two objects at a time.  Each pair is compared, and a decision is made based on which of the two objects is preferred. This methodology addresses the problem of determining the scale values of a set of objects on a preference continuum that is not directly observable.

Let $\pi_{(ij)i}$ denote the probability that the object $i$ is preferred in comparison with $j$. The probability that $j$ is preferred is $\pi_{(ij)j} = 1-\pi_{(ij)i}$. The basic Bradley-Terry (BT) model can be defined as a quasi-symmetry model for paired comparisons \citep[][p. 436]{agresti2002}

\begin{equation}
\label{eq1}
    \pi_{(ij)i} = \frac{\pi_i}{\pi_i + \pi_j},
\end{equation}
where $\pi_i$ and $\pi_j$ are non-negative parameters (also called worth parameters) describing the location of objects on the preference scale. 

The BT model can be expressed as a logistic model for paired preference data. Suppose to have a set of $n_o$ objects to be judged. The BT model has object parameters $\lambda_i^O$ such that 

\begin{equation}
\label{eq2}
    logit(\pi_{(ij)i}) = \log\left(\frac{\pi_{(ij)i}}{\pi_{(ij)j}}\right) = \lambda_i^O - \lambda_j^O,
\end{equation}

where $\lambda_i^O$ and $\lambda_j^O$ are object parameters related to $\pi$'s in Equation (\ref{eq1}) by $\lambda_i^O = \frac{1}{2}\ln(\pi_i)$.  The superscript $O$ refers to object-specific parameters. Thus, $\hat{\pi}_{(ij)i} = \frac{\exp{(\hat{\lambda_i}^O - \hat{\lambda_j}^O)}}{1+\exp{(\hat{\lambda_i}^O - \hat{\lambda_j}^O)}} $, where $\pi_{(ij)i} = \frac{1}{2}$ when $\lambda_i^O = \lambda_j^O$.  The model estimates  $\binom{n_o}{2}$ probabilities, which is the number of paired comparisons with $n_o$ objects. Note that the logit model in Equation (\ref{eq2}) is equivalent to the quasi-symmetry model in Equation (\ref{eq1}). In addition, identifiability of these two formulation requires a restriction on the parameters related on the last object $n_o$ such as $\lambda_{n_o}^O = 0$ or $\sum_i^{n_o} \pi_i$ = 1  \\

For each pair $i\geq j$, let $n_{ij}$ be the number of comparisons made between object $i$ and $j$, $y_{(ij)i}$ denotes the number of preferences of $i$ to $j$ and $y_{(ij)j}=n_{ij}-y_{(ij)i}$ denotes the number of preferences of $j$ to $i$. Assuming that $n_{ij}$ comparisons are independent and have the same probability $\pi_{(ij)i}$, the $y_{(ij)i}$ are binomially distributed with parameters $n_{ij}$ and $\pi_{(ij)i}$. 

The Bradley-Terry model can also be fitted as a log-linear model \citep{fienberg1976, sinclair1982, dittrich1998}. Among these authors,
\cite{sinclair1982} introduced a different approach: in comparing object $i$ with object $j$, the random variables $y_{(ij)i}$ and $y_{(ij)j}$ are assumed to follow a Poisson distribution.

Let $m(y_{(ij)i})$ be the expected number of comparisons in which $i$ is preferred to $j$. Then, using the respecification proposed by Sinclair and the notation for log-linear models for contingency tables, $m(y_{(ij)i}) = n_{ij}\pi_{(ij)i}$ has a log-linear representation

\begin{align}
\begin{aligned}
    \ln(m(y_{(ij)i})) = \mu_{ij} + \lambda_i^O - \lambda_j^O \\
    \ln(m(y_{(ij)j})) = \mu_{ij} - \lambda_i^O + \lambda_j^O,
\end{aligned}\label{eq:state-space&obs-equ}
\end{align}
where the nuisance parameters $\mu$ are defined by

\begin{equation}
\label{eq4}
    \mu_{ij} =  n_{ij} - \ln \left(\sqrt{\frac{\pi_i}{\pi_j}}+ \sqrt{\frac{\pi_j}{\pi_i}}\right),
\end{equation}

\noindent and they can be interpreted as interaction parameters representing the objects involved in the respective comparison, therefore fixing the corresponding $n_{ij}$ marginal distributions. In total, $2\binom{n_o}{2}$ expected counts are estimated. 

This approach allows synthesizing the information about all preferences in a unique design matrix. The design matrix is composed by column vectors representing the responses $y_{(ij)}$, the nuisance parameters $\mu_{ij}$, and the object parameters $\lambda_i^O$. For example, given three objects (A B C), an example of a design matrix is given in Table \ref{table1}.

\begin{table}[h!]
\centering
\caption{Design matrix with one judge and three objects: The first column indicates if the object $i$ is preferred ($y_{ij} = 1$) or not ($y_{ij} = 0$) in a certain preference for each pair of objects $ij$. The second column serves as an index for the $n \times (n-1) / 2$ comparisons. Finally, preferences are expressed in the last three columns. For example, the first line shows that object $B$ is preferred to $A$ since $y_{ij} = 1$, $\lambda_B^O = 1$, and $\lambda_A^O = -1$.}
\begin{tabular}{crrrr}
 & & & & \\
\hline
{$Response$} & \multicolumn{1}{c}{{$\mu$}} & \multicolumn{1}{c}{{$\lambda_A^O$}} & \multicolumn{1}{c}{{$\lambda_B^O$}} & \multicolumn{1}{c}{{$\lambda_C^O$}} \\ \hline
{$y_{AB} = 1$} & 1 & -1 & 1 & 0 \\
{$y_{AB} = 0$} & 1 & 1 & -1 & 0 \\
{$y_{AC} = 1$} & 2 & -1 & 0 & 1 \\
{$y_{AC} = 0$} & 2 & 1 & 0 & -1 \\
{$y_{BC} = 1$} & 3 & 0 & 1 & -1 \\
{$y_{BC} = 0$} & 3 & 0 & -1 & 1 \\ \hline
\end{tabular}
\label{table1}
\end{table}

The following equation gives the linear predictor $\eta$ for the basic LLBT model \citep{hatzinger2012}

\begin{equation}
\label{eq6}
    \eta_{y_{(ij)i}} = \ln(m(y_{(ij)i})) = \mu_{ij} + y_{(ij)i}(\lambda_i^O - \lambda_j^O).
\end{equation}

The log-linear formulation allows extending the model with multiple subject-specific  covariates. 

\subsection{The extended Bradley-Terry model with subject-specific covariates} 

In some cases, it could be interesting to analyze the variation of preferences according to subject-specific characteristics. The Bradley-Terry model can be extended to incorporate categorical or continuous covariates.

For a categorical covariate $S$, let $m(y_{(ij)i,l})$ be the expected number of preferences for $i$ compared with $j$, among individuals classified in covariate category $l$, with $l = 1 … L$, where $L$ represents the total number of levels of the covariate. The Bradley-Terry model is then specified as

\begin{align}
\begin{aligned}
    \ln  (m\left(y_{(ij)i,l}\right)) = \mu_{ij,l} + \lambda_i^O - \lambda_j^O + \lambda_l^S + \lambda_{i,l}^{OS} - \lambda_{j,l}^{OS} \\
     \ln  (m\left(y_{(ij)j,l}\right)) = \mu_{ij,l} - \lambda_i^O + \lambda_j^O + \lambda_l^S - \lambda_{i,l}^{OS} + \lambda_{j,l}^{OS}.
\end{aligned}\label{eq:state-space&obs-equ}
\end{align}
The parameter $\lambda_l^S$ represents the main effect of the subject-specific covariate $S$ measured on its $l$-th level; $\lambda_{i,l}^{OS}$ and $\lambda_{j,l}^{OS}$ are the subject-object interaction parameters describing the effect of $S$ observed on category $l$ and concerning the preference for object $i$ and $j$, respectively. 
The model parameters of interest $\lambda_{i,l}^{OS}$ and $\lambda_{j,l}^{OS}$ can again be interpreted in terms of log-odds and as a log-odds ratio 

\begin{equation}
\label{eq8}
    \ln\left(\frac{\pi_{(ij)i,l}}{\pi_{(ij)j,l}}\right) = 2(\lambda_i^O + \lambda_{il}^{OS}) - 2(\lambda_j^O + \lambda_{jl}^{OS}).
\end{equation}
If the covariate $S$ has no effect on the preferences of the judges, then $\lambda_{i,l}^{OS} = 0$. It means that the model collapses into the previously described basic BT model, and there is just one log-odds for the comparison of two specific objects. However, if there is a covariate effect so that there is at least one interaction parameter between the individuals and the subject-specific covariate that is significantly different from 0, we must distinguish different log-odds for each comparison and each significant subject-object interaction parameter \citep{hatzinger2012}.

When continuous subject-specific covariates are included, it is necessary to build up a separate contingency table for each judge, and each different value of the covariate. Hence, the LLBT equation for the $h$-th judge and objects $i$ and $j$ is

\begin{equation}
\label{eq9}
    \ln  (m\left(y_{(ij)i,h}\right)) = \mu_{ij,h} + y_{(ij)i,h}(\lambda_{i,h}^O-\lambda_{j,h}^O).
\end{equation}

The parameter $\lambda_{i,h}^O$ can be expressed  through a linear relation 

\begin{equation}
\label{eq10}
\lambda_{i,h}^O = \lambda_i^O + \sum_{p = 1}^P \beta_{ip}x_{p,h},
\end{equation}

where $x_{p,h}$ corresponds to the value of the $x_p$-th continuous covariate $(p = 1...P)$ observed for judge $h$. The parameters $\beta$ can be interpreted as the effect of the covariates on object $i$, whilst  $\lambda_i^O$ acts as intercept and indicates the location of object $i$ in the overall consensus ranking. 

Following this approach, it is possible to compute the deviance of the model as the deviance of a fitted Poisson regression

\begin{equation}
\label{eq11}
    D=2\sum_{h=1}^Hy_{ij,h}\times \log\left(\frac{y_{ij,h}}{\hat{y}_{ij,h}}\right),
\end{equation}

where $y_{ij,h}$ represents the observed values of each comparison $ij$ for each judge $h$, and $\hat{y}_{ij,h}$ are the predicted values based on the estimated model parameters.
This measure indicates how well the model fits the data. If the model fits well, the $y_{ij,h}$ will be close to their predicted values $\hat{y}_{ij,h}$.

\section{STIMA and trunk modeling}
\label{RT}
The Bradley-Terry model can be applied to preference data by specifying a regression model for paired comparisons. In this paper, this specification is aimed at estimating in an automatic and data-driven mode the main effects part of the model as well as, if present, its interaction effects part. For this purpose, we resort to the STIMA framework extended with the use of GLM in \cite{CD17}, and combine the extended Bradley-Terry model including subject-specific covariates with the regression trunk methodology \citep{dusseldorp2004}. The main feature of a regression trunk is that it allows the user to evaluate in a unique model and simultaneously the importance of both main and interaction effects obtained by first growing a regression trunk and then by pruning it back to avoid overfitting. The interaction effects are hereby intended as a particular kind of non-additivity which occurs if the individual effects of two or more variables do not combine additively \citep{bdg2007} or when over and above any additive combination of their separate effects, these variables have a joint effect \citep[][p. 257]{cohen2013}.

The implementation of STIMA is based on the integration between generalized linear models - GLM \citep{MN1989} and Classification And Regression Trees (CART) \citep{breiman1984}. A binary splitting algorithm with an ad-hoc defined splitting criterion and a stopping rule is used to model interaction terms in GLM. The estimated model including main effects and threshold interactions is equivalent, in its form, to a standard GLM with both random and systematic components and a link function.
Usually, this model is used when the analyst has no exact a priori hypotheses about the nature of the interaction effects. For example, regression trunks have been successfully applied in the framework of tourism website evaluation \citep{CCM19}. 

STIMA allows overcoming the problems related to both the additive nature of regression models and the lack of main effects in tree-based methods. Typically, regression models are hard to interpret when higher-order interactions are arbitrarily included. In contrast, CART-like decision trees quickly identify complex interactive structures but, when data includes also linear main effects,  they "would take many fortuitous splits to recreate the structure, and the data analyst would be hard-pressed to recognize them in the estimated tree" \citep[][p. 313]{hastie2009}.

Notationally, the generalized linear model estimated by STIMA assumes that a response variable $y$ observed on $n$ subjects has an exponential family density $\rho_y(y;\theta;\phi)$ with a natural parameter $\theta$ and a scale parameter $\phi$. The response $y$ depends on a set of $P$ categorical and/or continuous covariates $x_p$ ($p=1,\ldots,P)$ and its mean $\mu = E(y|x_1,\ldots,x_P)$ is linked to the $x_p$s via a link function $g(\cdot)$:

\begin{equation}
\label{eq12}
    g(\mu) = \eta = \beta_0 + \sum_{p=1}^P \beta_p x_{p,h} + \sum_{t = 1}^{T-1} \beta_{P+t} I\{(x_{1,h},\ldots,x_{P,h}) \in t \}
\end{equation}

Equation (\ref{eq12}) refers to a standard GLM presenting a linear predictor $\eta$ such that $\mu = g^{-1}(\eta)$ ($\mu$ is an invertible and smooth function of $\eta$). The first $P$ parameters concern the main effects part of the model estimated in the root node of the trunk via standard GLM, whilst the other $T-1$ parameters define the interaction effects part of the model obtained by partitioning recursively in a binary way the $n$ cases in order to add additional interaction terms defined by the coefficients $\beta_{P+t }$ and the indicator variables $I\{(x_{1,h},\ldots,x_{P,h}) \in t \}$. Since a tree structure with $T$ terminal nodes is derived recursively, the so-called trunk, $I\{(x_{1,h},\ldots,x_{P,h}) \in t \}$ with ($t=1,\ldots,T-1$) refers to the subset of cases belonging to the terminal node $t$ of the trunk. The interaction effect of the $T$-th terminal node is not considered as this node serves as reference category for the other interaction effects. Being obtained by a sequential binary splitting of the original data, the interaction effects correspond to threshold interactions since the values/labels of the splitting predictors leading to a specific terminal node can be considered as thresholds that partition the predictor space in order to correctly identify a GLM with interaction effects that maximizes goodness of fit by controlling for overfitting. 

In a generic iteration of STIMA, adding a new threshold interaction effect in the model means adding a new binary split to the trunk. This happens when the candidate split maximizes the effect size of the model. The search of the additional interaction effect is conducted by considering for each predictor $x_p$ all possible split points for each current terminal node. An additional interaction effect is included if the effect size between the model estimated before the current split and that including the candidate interaction originating from the current split is maximized. Once the split is found, all regression coefficients in the model are re-estimated.\\ In the case of a continuous response, $g(\cdot)$ corresponds to the identity function and the effect size is computed as the relative increase in variance-accounted-for. The resulting model is the standard regression trunk model \citep{dusseldorp2010combining}. Whereas, if one assumes that observations are independent realizations of Binomial random variables the link function corresponds to the Logit function and the effect size is computed as the relative increase in the log-likelihood $R^2$ observed when passing from the model which does not include the candidate interaction effect to the one that includes it. The resulting model is the logistic classification trunk \citep{CD17}. 

In all cases, STIMA works by first growing a full trunk, corresponding to the maximum number of splits $T-1$, and then pruning it back using $V$-fold cross-validation with the $c$ standard error rule ($c \cdot SE$ rule). The constant $c$ varies between 0 and 1, and the higher its value the more the tree is pruned back.  

\section{The Bradley-Terry Regression Trunk (BTRT) for preference data}
\label{prop}
In the following, we introduce the Bradley-Terry Regression Trunk (BTRT) model for the analysis of preference data. It combines the extended log-linear Bradley-Terry model including subject-specific covariates introduced in Equations \ref{eq9} and \ref{eq10} with the STIMA-based trunk model specified in Equation \ref{eq12}. The resulting model is still a log-linear model aimed at modeling the pairwise comparisons of objects $i$ and $j$ (Equation \ref{eq9}) through a different specification of the linear components describing the consensus expressed for the objects (see for example Equation \ref{eq10} for object $i$). In particular, using the regression trunk approach and considering the possible effect of subject-specific covariates $x_p$ the estimated consensus expressed for object $i$ by the judge $h$ is 

\begin{equation}
\label{eq13}
    \hat \lambda_{i,h} = \hat \lambda_i + \sum_{p=1}^P \hat \beta_{i,p} x_{p,h} + \sum_{t = 1}^{T-1} \hat \beta_{i,P+t} I\{(x_{1,h},\ldots,x_{P,h}) \in t \}
\end{equation}

Again, the term $\sum_{p=1}^P \hat \beta_{i,p} x_{p,h}$ is the main effects part assessing the effects of covariates on the consensus for object $i$. The interaction effects part is estimated by $\sum_{t = 1}^{T-1} \hat \beta_{i,P+t} I\{(x_{1,h},\ldots,x_{P,h}) \in t \}$ and is derived from the terminal nodes of a regression trunk that searches for possible threshold interactions between the $P$ covariates assuming they have a joint effect on the consensus expressed for object $i$ besides their individual (main) effect. Thus, the regression trunk has $T$ terminal nodes and for each terminal node $t$ an additional parameter $\beta_{i,P+t}$ is estimated. It expresses the effect of the threshold interaction between the covariates $x_1,\ldots,x_P$ whose split points lead to $t$. The estimated intercept term $\hat \lambda_i$ measures the average consensus about object $i$ in the root node of the trunk whilst the estimated intercept for the terminal node $t$ is $\hat{\lambda}_i + \hat{\beta}_{i, P+t}$. Note that the subscript $O$ is left out from the notation of the    $\hat{\lambda}$ parameters for readability reasons.

Basically, the estimation procedure of BTRT  is framed within the STIMA algorithm, but some steps are different. Once a set of paired comparisons is given, a preliminary data processing step is required to obtain the design matrix of the Bradley-Terry model. In our framework, ties are not allowed. The final design matrix is composed of $n=n_o \times (n_o-1) \times H$ rows, where $H$ indicates the number of judges. The total number of rows is equal to the product between the number of comparing objects, that is 2, the number of paired comparisons ($n_o \times (n_o-1)/2)$, and the number of judges, resulting in $2 \times (n_o \times (n_o-1)/2) \times H$.

In the above-described framework, estimating a BTRT model needs three basic ingredients: a splitting criterion, a stopping rule, and a pruning procedure.

\subsection{Growing the trunk}\label{growing}

In each step of STIMA, a generalized linear model with a Poisson link is fitted to the data. To discover the main effects, it is only necessary to fit the model in the root node. The first estimated model consists of $P$ coefficients $\beta$ that describe the probability distribution of preferring a particular object to another one, given a set $(x_1,...,x_P)$ of judges' characteristics. STIMA searches for a split among all the values for each continuous covariate. In each step of the regression trunk building procedure, splitting a parent node means finding a dichotomous variable $z^*_{ijp,t}$ that updates the indicator function $I(\cdot)$ introduced in Equation (\ref{eq13}). For each terminal node $t$ of the trunk, the number of dichotomous variables $z^*_{ijp,t}$ is equal to the number of splits leading to $t$. The interaction effects part of Equation (\ref{eq13}) contains $T-1$ terms since one terminal node is treated as the reference group.

The search of the best split of the trunk at each iteration is made by taking into account all the available terminal nodes at that step. For a particular terminal node and based on paired comparisons, for each covariate $x_p$, with $(p=1,\ldots P)$, we consider each unique value of $x_{p}$ as a candidate split point. Specifically, a Bradley-Terry model is estimated for each of the possible pairs of candidate values $ij \in [1,n_o]; i \neq j$, by discretizing $x_p$ and creating the associated dichotomous variable $z_{ijp}$. 

Next, the split point associated with $z^*_{ijp}$ maximizing the decrease in deviance is computed with respect to the goodness-of-fit test based on the deviance of a Poisson regression model introduced in Equation (\ref{eq11}). Thus, it is considered as the "best" split point and the node is split according to the specific value of the discretized variable $x_p$. The splitting criterion of BTRT is based on maximizing the decrease in deviance when moving from a parent node to the two possible daughter nodes defined by splitting on $z_{ijp}$. This is equivalent to comparing the fit of two nested models, one simpler and one more complex, and could lead to a profile log-likelihood ratio test of the hypothesis that the extra parameter $\beta_{P+t}$ is zero. 

This split search procedure is repeated by searching for each splitting node $t$ the best split point so that, once found, the new dichotomous variable $z^*_{ijp,t}$ is added to the model and an additional interaction effect is included. When the split is found, all regression coefficients in the model are re-estimated. 

Preliminarily, the user is required to choose between two main approaches that could be followed in BTRT:\\ 
a) {\it One Split Only (OSO)}, where the splitting covariates already used in the previous splits are not considered as candidate splitting variable for the current split;\\ 
b) {\it Multiple Splitting (MS)}, where the whole set of covariates is considered to split the current node despite some of them have been previously selected to split other nodes.\\ 
The OSO approach returns a tree in which it is possible to analyze the interaction effects between all the covariates. In this case, the final tree might not necessarily return the best model as that producing the best goodness of fit (i.e., maximum reduction in deviance). Besides, following the MS approach it is possible to achieve the maximum reduction in deviance, but there is a risk of obtaining a tree that utilizes the same covariate (with different values) to split several, even subsequent, nodes. In this case, it can happen that only the main effects part is retained and thus it is not possible to analyze interactions. We compare the two criteria in the real data application (see Section \ref{RD}).

At each split step, the estimated regression parameters $\hat \beta_{i,P+t}$ measure the probability of preferring a specific object $i$, given the interaction between different characteristics of a particular group of judges.
While some similar methods, such as M5 \citep{Q92} and Treed regression \citep{AG96}, estimate several linear models, one in each node of the tree, the regression trunk model estimates a single linear model only. 

\begin{figure}[h!]
\centering
\resizebox{\textwidth}{!}{%
\begin{tikzpicture}[align=center, auto]
{\scriptsize
  \node[block]			(block1)  {Estimation of the main effects model in the root node of the trunk:\\
	$ \hat \lambda_{i,h} = \hat \lambda_i + \sum_{p=1}^P \hat \beta_{i,p}x_{p,h} $};
  \node[block, below of=block1, node distance=1.8cm]     (block2)     {Split search at current node $t_c$: for each value of each continuous subject-specific covariate $x_p$ find the dichotomous variable $z_{ijp,t}^*$ that minimizes the loglikelihood deviance of the model};
  \node[block, below of=block2, node distance=1.8cm]     (block3)   {Does $z_{ijp,t}^*$ cause a significant decrease in model deviance?};
  \node[blockyes, below of=block3, node distance=1.2cm]  (block4)   {Yes};
  \node[block, below of=block4, node distance=1.2cm]  (block5)    {Create child nodes $t_{c+1}$ and $t_{c+2}$};
  \node[block, below of=block5, node distance=1.8cm]   (block6)    {Is the number of cases in both $t_{c+1}$ and $t_{c+2}$ greater or equal to the square root of the number of cases in $t_c$?};
	\node[blockyes, below of=block6, node distance=1.1cm]   (block7)   {Yes};
	\node[block, below of=block7, node distance=1.7cm]   (block8)     {$z_{ijp,t}^*$ updates the indicator function $I\{(x_{1,h},\ldots,x_{P,h}) \in t$ and the model including the last threshold interaction effect is re-estimated:
$$\hat \lambda_{i,h} = \hat \lambda_i+\sum_{i=1}^P \hat \beta_{i,p}x_{p,h}+\sum_{t = 1}^{T-1} \hat \beta_{i,P+t} I\{(x_{1,h},\ldots,x_{P,h}) \in t \}$$};
  \node[cloud, right of=block3, node distance=6cm](block9)  {No};      
	\node[cloud, right of=block6, node distance=6cm]  (block10)  {No};
	\node[decision, right of=block5, node distance=8cm]  (block11)  {Stop};   
  \draw[->]             (block1) -- (block2);
  \draw[->]     				(block2) -- (block3);
  \draw[->]      				(block3) -- (block4);
	\draw[->]             (block4) -- (block5);
  \draw[->]     				(block5) -- (block6);
  \draw[->]      				(block6) -- (block7);
	\draw[->]      				(block7) -- (block8);
  \draw[->]     (block3) -- (block9);
	\draw[->]     (block6) -- (block10);
	\draw[->]     (block9) -- (block11);
	\draw[->]     (block10) -- (block11);
	\draw[->]     (block8) -- +(-6,0) |- (block2);  
	}
  \end{tikzpicture}
	}
\caption{Flowchart of the STIMA algorithm implementing the BTRT model for preference data}
\label{fig1}
\end{figure}
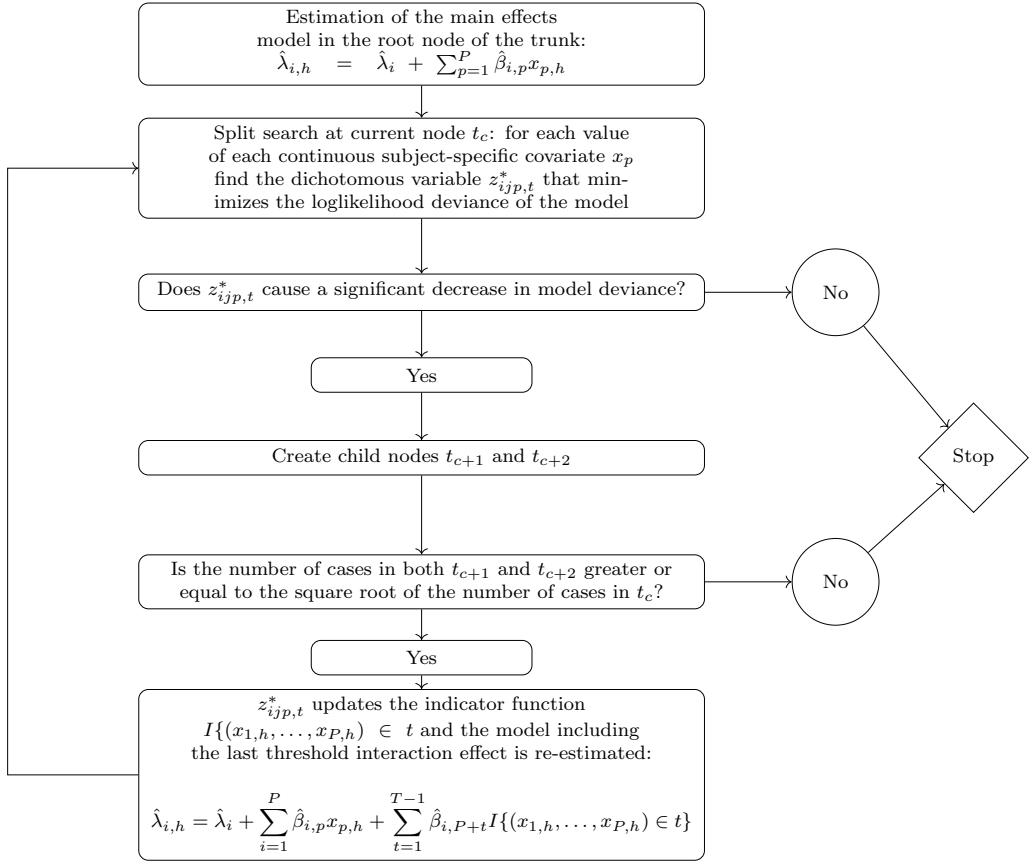

Consistent with standard criteria applied in decision tree modeling, the stopping criterion of BTRT is based on the a-priori definition of the minimum number of observations for a node to be split. The default implementation is based on the requirement that the size of the splitting node should be at least equal to the square root of the size of its parent node and, in any case, the splitting node should include more than $4$ observations.
Figure \ref{fig1} shows a flowchart in which the tree growing procedure is schematically explained.

\subsection{Pruning the trunk}\label{pruning}

When the final estimated trunk model presents a large number of higher-order interactions it may be challenging to interpret the results and the overfitting problem might occur. Anyway, growing the maximum expanded trunk is necessary since a small trunk may not be able to capture the real interactive structure of the data if the splitting process ends too early. For this reason, BTRT considers a pruning procedure operated after the trunk growing. In particular, a $V$-fold cross-validation of the BTRT model deviance is computed for each step split of the trunk. The user has to provide the number of subsets $V$ in which the entire data set is divided. To obtain the cross-validated deviance, all the preferences expressed by a particular judge $h$ in the design matrix are randomly assigned to a specific subset and, for $V$ times, the BTRT trunk model estimated in a specific node is trained on $V-1$ subsets whilst the left-out subset is treated as a test set. At the end of the process, a predicted value ${\hat{y}_{ij,h}}$ is obtained for each observation in the data matrix. Following this approach, the case-wise cross-validation deviance $D^{cv}$ is

\begin{equation}
\label{eq:state-space&obs-equ}
    D^{cv} =\frac 1 n \left [2 \sum_{i'=1}^n  y_{i'j;h}\times \log \left(\frac{y_{i'j;h}}{\hat{y}_{i'j;h}}\right) \right ], \hspace{1cm} (i',j) \in n_o,  (i'\neq j), 
h \in H    
\end{equation}

where $n$ is equal to the total number of rows of the design matrix and $i'$ is its generic row. Note that the number of rows $n$ is greater than the total number of judges $H$.  The standard error of $D^{cv}$ is 
\begin{equation}
\label{eq15}
SE^{cv} = \sqrt{ \frac{1}{n}  \sum_{i'=1}^n   \left [ y_{i'j;h} \times \log \left(\frac{y_{i'j;h}}{\hat{y}_{i'j;h}}\right) - D^{cv} \right ]^2}
\end{equation}

Usually, $D^{cv}$ decreases after the first splits of the trunk and starts to increase next. BTRT uses the same $c \cdot SE$ pruning rule used in STIMA. 
Let $t^* \in [1,T]$ be the size of the regression trunk with the lowest $D^{cv}$, say $D^{cv}_{t^*}$. The best size of the BTRT trunk $t^{**}$ corresponds to the minimum value of $t$ such that $D^{cv}_{t^{**}} \leq D^{cv}_{t^*} + c \cdot SE^{cv}_{t^*}$. We investigate about the optimal choice of the pruning parameter $c$ in Section \ref{sim}.

\section{Simulation study: the choice of the pruning parameter}
\label{sim}

Pruning the BTRT model with the $c$ $\cdot$ SE rule requires the choice of the most suitable value for the parameter $c$. The optimal value may depend on characteristics of the data, such as sample size (Dusseldorp et al., 2010). In this section, a simulation study is carried out to assess the value of the optimal $c$ to be used to select the final BTRT model. 

For the regression trunk approach used to detect threshold interactions in the linear model, \cite{dusseldorp2010combining} reported that most of the times a value of $c = 0$ results in a regression trunk with too many interaction terms whilst a value of $c = 1$ gives a small-sized regression trunk with too few interaction terms. 

As for BTRT, we compare the performance of seven pruning rules obtained by specifying seven different values of $c$ ranging from $0$ to $1$, namely: 0.00, 0.10. 0.30, 0.50, 0.70, 0.90 and 1.00. 

Three different scenarios are considered for the data generating process (DGP): 
\begin{equation}
\label{eq16}
    \lambda_{i,h} = \lambda_i + \beta_{i,1}x_{1,h};
\end{equation}

\begin{equation}
\label{eq17}
    \lambda_{i,h} = \lambda_i + \sum_{p=1}^4 \beta_{i,p}x_{p,h};
\end{equation}

\begin{equation}
\label{eq18}
    \lambda_{i,h} = \lambda_i + \sum_{p=1}^4 \beta_{i,p}x_{p,h} + \beta_{i,5}I(x_{1,h} > 0.00 \cap x_{2,h} > 0.50).
\end{equation}

In the first scenario (Equation \ref{eq16}), only one subject-specific covariate ($x_1$) affects the preferences expressed by the generic judge $h$ on each object $i$. In the second one (Equation \ref{eq17}), four subject-specific covariates are assumed to influence the judges' preferences. These two models present  linear main effects only so that the performance metric of the pruning rules is  the proportion of times a BTRT model with at least one interaction term is selected (Type I Error). In the third scenario (Equation \ref{eq18}) a model including both linear main effects and threshold interaction effects is considered as a threshold interaction term between $x_1$ and $x_2$ is added to the main effects part of the model. In this case, the performance metric of the pruning rule is the Type II Error, obtained by computing the proportion of times the selected regression trunk model does not include $x_1$ and $x_2$ exactly as the first and only two interacting variables. In all cases, all the covariates $x_p$ are standard normally distributed. 

\subsection{Design factors and procedure}

Three design factors are considered in the simulation study:

\begin{itemize}
    \item The number of judges $H$: 100, 200, 300;
    \item The number of objects $n_o$: 4, 5. The consensus rankings were set as (A B C D) and (A B C D E), respectively, by using decreasing values of $\lambda_i$, namely $(0.9, 0.4, 0.3, 0.0)$ in the first case, and $(0.8, 0.4, 0.2, 0.0, 0.1)$ in the second one;    
    \item The effect size of each covariate $x_p$ on the preferences expressed by the judge $h$ on each object $i$. Values of the parameters $\beta_i$ are reported in Table \ref{table2} for each set of objects, the two possible effect sizes and the three different scenarios.    
\end{itemize}

\begin{table}[!h!]
\caption{Simulated values of $\beta_i$ for the estimation of the pruning parameter $c$}
\resizebox{\textwidth}{!}{%
\begin{tabular}{lcccccccccccc}
\hline
 N. objects = 4 & & & & & &  & & & & & &\\
Effect-size                                 &      \multicolumn{4}{c}{Low}& & \multicolumn{6}{c}{High}  \\
object                                        & A    &  B   &  C   &  D   & &        & A   &  B   &  C   &  D   & \\
\hline & & & & &  & & & & & & & \\
 & \multicolumn{11}{c}{1st scenario (Equation \ref{eq16})}\\
$\beta_1$                             & 0.30 & 0.20 & 0.10 & 0.00 &  &        & 0.90& 0.80 & 0.70 & 0.00 &    \\ 
  & & & & &   & & & & & & & \\
& \multicolumn{11}{c}{2nd scenario (Equation \ref{eq17}):  add $\beta_2$, $\beta_3$ and $\beta_4$}\\ 
$\beta_2$                             & 0.20 & 0.30 & 0.10 & 0.00 &   &       & 0.80 & 0.70 & 0.90 & 0.00& \\  
$\beta_3$                             & 0.10 & 0.20 & 0.30 & 0.00 &    &      & 0.70 & 0.90 & 0.80 & 0.00&  \\
$\beta_4$                             & 0.30 & 0.10 & 0.20 & 0.00 &     &     & 0.90 & 0.70 & 0.80 & 0.00& \\
  & & & & &   & & & & & & & \\
& \multicolumn{11}{c}{3rd scenario (Equation \ref{eq18}):  add $\beta_5$}\\ 
$\beta_5$                             & 0.25 & 0.15 & 0.35 & 0.00 & &         & 0.55 & 0.65 & 0.45 & 0.0  &  \\ \hline
  & & & & &   & & & & & & & \\ 
N. objects = 5 & & & & & &  & & & & & &\\
Effect-size                                 &      \multicolumn{4}{c}{Low}& & \multicolumn{6}{c}{High}  \\
object                                        & A    &  B   &  C   &  D   &   E    &  & A   &  B   &  C   &  D   & E\\
 \hline & & & & &  & & & & & & & \\
 & \multicolumn{11}{c}{1st scenario (Equation \ref{eq16})}\\
$\beta_1$                             & 0.40 & 0.30 & 0.20 & 0.10 & 0.00 & &       0.90 & 0.80 & 0.70 & 0.60 & 0.00\\ 
  & & & & &   & & & & & & & \\
& \multicolumn{11}{c}{2nd scenario (Equation \ref{eq17}):  add $\beta_2$, $\beta_3$ and $\beta_4$}\\ 
  & & & & &   & & & & & & & \\
$\beta_2$ & 0.30 & 0.20 & 0.10 & 0.40 & 0.00 & &         0.80 & 0.90 & 0.60 & 0.70 & 0.00 \\  
$\beta_3$ & 0.20 & 0.10 & 0.30 & 0.40 & 0.00 &  &        0.70 & 0.60 & 0.80 & 0.90 & 0.00 \\
$\beta_4$ & 0.10 & 0.20 & 0.40 & 0.30 & 0.00 &  &        0.90 & 0.70 & 0.60 & 0.80 & 0.00  \\
  & & & & &   & & & & & & & \\
& \multicolumn{11}{c}{3rd scenario (Equation \ref{eq18}):  add $\beta_5$}\\ 
$\beta_5$                             & 0.25 & 0.15 & 0.35 & 0.45 & 0.00       &    & 0.55 & 0.65 & 0.45 & 0.60 & 0.00\\ \hline \hline
\end{tabular}
}
\label{table2}
\end{table}

The combination of these three design factors ($n_o \times H \times$ {\it effect size}) results in 12 different BTRT specifications. For each of them, we generate 100 random samples, so that 1,200 data sets were generated for each true scenario, given in Equations (\ref{eq16}), (\ref{eq17}), and (\ref{eq18}). In each run, a BTRT with a maximum of five terminal nodes ($T$ = 5) is estimated.   

Once the design factors are set,  following Equation \ref{eq1} the values of $\hat{\lambda}_{i,h}$  are estimated in order to obtain the probability that a  judge $h$ prefers the object $i$ to $j$. The latter are computed for each possible comparison as follows      

\begin{equation}
\label{eq19}
        \pi_{(ij)i,h} = \frac{\exp{[2(\hat{\lambda}_{i,h}-\hat{\lambda}_{j,h})]}}{1+\exp{[2(\hat{\lambda}_{i,h}-\hat{\lambda}_{j,h}})]};
     \end{equation}
    
The design matrix of the log-linear Bradley Terry model requires the values of $y$ in the first column. The response $y$ is coded as a 0-1 variable depending on whether or not an individual preference occurs for each comparison $ij$. Thus, we consider $y_{ij,h}$ as the realization of a Bernoulli distribution that assumes the value $1$ with probability $\pi_{(ij)i,h}$. The main problem for this kind of coding is that it is possible to obtain combinations of 0-1 values for the same judge that do not verify the transitivity property between the preferences. The number of all possible combinations of two values for each judge is equal to $2^{\frac{n_o(n_o-1)}{2}}$, where the exponent is the number of paired comparisons obtainable from $n_o$ objects. However, when ties are not allowed, the number of permutations of $n_o$ objects is equal to $n_o!$, which is much smaller than the number of all the possible combinations of two values. When $n_o$ is higher then 3, it is very likely to obtain combinations that do not find a counterpart in the universe of allowed rankings. To avoid this problem, we replaced the combinations not allowed with the closest permutation in the universe of $n_o!$ rankings. 

\subsection{Results}
Results of the simulation study are summarized in Tables \ref{table3}, \ref{table4} and \ref{table5}. For the first two scenarios, the pruning rules are evaluated with respect to the Type I error (Tables \ref{table3}, \ref{table4}) whilst for the third scenario the focus is on the Type II error (Table \ref{table5}). To facilitate the interpretation of the results, the tables for Type II error show the power of the pruning rules (i.e., 1 - error), rather than the Type II errors. Results are reported for the $9$ different values of the $c$ parameter (0, 0.1, 0.3, 0.5, 0.7, 0.9, 1), as well as for the number of objects (4 or 5), the number of judges (100, 200 or 300) and the effect sizes (Low or High). A threshold value of $0.05$ is used for Type I error so that higher values are shown in boldface because the error is too high. For power we used the value $0.8$ as threshold so that if the power is less than $0.8$, then the power is too small and the values are shown in boldface.  

Table \ref{table3} reports the results for the first scenario where only the main effects of the single covariate $x_1$ are considered. When the number of objects is equal to 4 and the effect of $x_1$ is low, the pruning rules with $c \geq 0.1$ result in acceptable Type I errors despite the sample size. However, when the effect size increases, the case with $H = 100$ requires higher values of $c$ (i.e., $c \geq 0.3$) for the pruning parameter. When the number of objects is equal to 5 the inverse situation is observed: for small effect sizes higher values of $c$ (i.e., $c \geq 0.5$) are required, whilst for a high effect sizes lower values of $c$ (i.e., $c \geq 0.3$) can be used.  

\begin{table}[h]
\centering
\caption{Results first scenario: Type I error. Error higher than 0.05 in boldface.}
\resizebox{\textwidth}{!}{%
\begin{tabular}{l|lllllll|llllllll} \toprule
N. objects        &      &      &      & $n_o = 4$ &      &      &      &  &      &      &      & $n_o = 5$ &      &      &      \\
Effect size        &      & Low  &      &       &      & High &      &  &      & Low  &      &       &      & High &      \\
N. judges        & 100  & 200  & 300  &       & 100  & 200  & 300  &  & 100  & 200  & 300  &       & 100  & 200  & 300  \\ \midrule
c = 0.0   & \textbf{0.76} & \textbf{0.82} & \textbf{0.82} &       & \textbf{0.95} & \textbf{1.00} & \textbf{1.00} &  & \textbf{0.80} & \textbf{0.90} & \textbf{0.98} &       & \textbf{0.75} & \textbf{0.84} & \textbf{0.82} \\
c = 0.1 & \textbf{0.16} & \textbf{0.18} & 0.04 &       & \textbf{0.62}  & \textbf{0.51} & \textbf{0.58}  &  & \textbf{0.60}  & \textbf{0.58} & \textbf{0.60} &       & \textbf{0.30} & \textbf{0.38} & \textbf{0.26} \\
c = 0.3 & 0.01 & 0.00 & 0.00 &       & \textbf{0.26} & \textbf{0.12} &\textbf{0.08} &  & \textbf{0.32} & \textbf{0.18} & \textbf{0.28} &       & \textbf{0.08} & \textbf{0.08} & 0.00 \\
c = 0.5 & 0.00 & 0.00 & 0.00 &       & \textbf{0.08} & 0.05 & 0.02 &  & \textbf{0.12} & 0.04 & \textbf{0.10} &       & 0.00 & 0.02 & 0.00 \\
c = 0.7 & 0.00 & 0.00 & 0.00 &       & 0.03 & 0.00 & 0.00 &  & 0.04 & 0.02 & 0.00 &       & 0.00 & 0.00 & 0.00 \\
c = 0.9 & 0.00 & 0.00 & 0.00 &       & 0.00 & 0.00 & 0.00 &  & 0.02 & 0.02 & 0.00 &       & 0.00 & 0.00 & 0.00 \\
c = 1.0   & 0.00 & 0.00 & 0.00 &       & 0.00 & 0.00 & 0.00 &  & 0.02 & 0.02 & 0.00 &       & 0.00 & 0.00 & 0.00 \\ \bottomrule

\end{tabular}%
}
\label{table3}
\end{table}

Table \ref{table4} displays the Type I errors when all the covariates $x_1,...,x_4$ influence judges' preferences individually (second scenario). In this case, for $n_o=4$ the values of $c \geq 0.3$ provide acceptable error rates despite the effect size. compared to the situation in which the effect size is high; for $n_o=5$ and high effect size it would be better to choose a pruning parameter $c \geq 0.5$. 

\begin{table}[h]
\centering
\caption{Results second scenario: Type I error. Error higher than 0.05 in boldface.}
\resizebox{\textwidth}{!}{%
\begin{tabular}{l|lllllll|llllllll} \toprule
N. objects        &      &      &      & $n_o = 4$ &      &      &      &  &      &      &      & $n_o = 5$ &      &      &      \\
Effect size        &      & Low  &      &       &      & High &      &  &      & Low  &      &       &      & High &      \\
N. judges        & 100  & 200  & 300  &       & 100  & 200  & 300  &  & 100  & 200  & 300  &       & 100  & 200  & 300  \\  \midrule
c = 0.0   & \textbf{0.88} & \textbf{0.86} & \textbf{0.98} &       & \textbf{0.95} & \textbf{0.94} & \textbf{0.98} &  & \textbf{0.97} & \textbf{1.00} & \textbf{0.98} &       & \textbf{0.91} & \textbf{0.96} & \textbf{1.00} \\
c = 0.1 & \textbf{0.58} & \textbf{0.56} & \textbf{0.66} &       & \textbf{0.67}  & \textbf{0.66} & \textbf{0.74}  &  & \textbf{0.74}  & \textbf{0.86} & \textbf{0.86} &       & \textbf{0.62} & \textbf{0.70} & \textbf{0.80} \\
c = 0.3 & \textbf{0.14} & \textbf{0.06} & \textbf{0.10} &       & \textbf{0.11} & 0.04 & \textbf{0.10} &  & \textbf{0.09} & \textbf{0.14} & \textbf{0.12} &       & \textbf{0.16} & \textbf{0.28} & \textbf{0.18} \\
c = 0.5 & 0.04 & 0.02 & 0.00 &       & 0.01 & 0.00 & 0.00 &  & 0.01 & 0.02 & 0.04 &       &\textbf{0.06} & \textbf{0.06} & 0.02 \\
c = 0.7 & 0.02 & 0.00 & 0.00 &       & 0.00 & 0.00 & 0.00 &  & 0.00 & 0.00 & 0.00 &       & 0.01 & 0.00 & 0.00 \\
c = 0.9 & 0.00 & 0.00 & 0.00 &       & 0.00 & 0.00 & 0.00 &  & 0.00 & 0.00 & 0.00 &       & 0.01 & 0.00 & 0.00 \\
c = 1.0   & 0.00 & 0.00 & 0.00 &       & 0.00 & 0.00 & 0.00 &  & 0.00 & 0.00 & 0.00 &       & 0.00 & 0.00 & 0.00 \\ \bottomrule

\end{tabular}%
}
\label{table4}
\end{table}

The third scenario reflects the case in which all the covariates $x_1,...,x_4$ have an influence on the expressed preferences, and the first two covariates interact with each other, as shown in Equation \ref{eq18}. The power (1 - Type II error) is displayed in Table \ref{table5} for each possible value of $c$. It emerges that for $n_o=4$ a value of $c\geq 0.3$ is considered as satisfactory  despite the effect size (except in case there are 100 judges and low effect size), whilst for the $n_o=5$ case with high effect size it is preferable to increase the value of $c$ up to 0.9.

\begin{table}[h]
\centering
\caption{Results third scenario: Test's power (1-Type II error). Power lower than 0.80 in boldface.}
\resizebox{\textwidth}{!}{%
\begin{tabular}{l|lllllll|llllllll} \toprule
N. objects        &      &      &      & $n_o = 4$ &      &      &      &  &      &      &      & $n_o = 5$ &      &      &      \\
Effect size        &      & Low  &      &       &      & High &      &  &      & Low  &      &       &      & High &      \\
N. judges        & 100  & 200  & 300  &       & 100  & 200  & 300  &  & 100  & 200  & 300  &       & 100  & 200  & 300  \\  \midrule
c = 0.0   & \textbf{0.00} & \textbf{0.00} & \textbf{0.00} &       & \textbf{0.03} & \textbf{0.02} & \textbf{0.01} &  & \textbf{0.02} & \textbf{0.00} & \textbf{0.01} &       & \textbf{0.00} & \textbf{0.00} & \textbf{0.02} \\
c = 0.1 & \textbf{0.45} & \textbf{0.52} & \textbf{0.28} &       & \textbf{0.30}  & \textbf{0.20} & 0.80  &  & \textbf{0.22}  & \textbf{0.06} & \textbf{0.01} &       & \textbf{0.28} & \textbf{0.12} & \textbf{0.02} \\
c = 0.3 & \textbf{0.79} & 0.94 & 0.84 &       & 0.84 & 0.84 & 0.99 &  & 0.82 & \textbf{0.52} & \textbf{0.46} &       & \textbf{0.74} & \textbf{0.28} & \textbf{0.14} \\
c = 0.5 & 0.99 & 0.99 & 0.99 &       & 0.92 & 0.94 & 0.98 &  & 0.96 & 0.96 & 0.88 &                                  & 0.98 & \textbf{0.44} & \textbf{0.24} \\
c = 0.7 & 1.00 & 1.00 & 1.00 &       & 0.96 & 0.98 & 1.00 &  & 1.00 & 1.00 & 1.00 &       & 0.98 & 0.80 & \textbf{0.56} \\
c = 0.9 & 1.00 & 1.00 & 1.00 &       & 1.00 & 1.00 & 1.00 &  & 1.00 & 1.00 & 1.00 &       & 1.00 & 1.00 & 0.90 \\
c = 1.0   & 1.00 & 1.00 & 1.00 &       & 1.00 & 0.98 & 1.00 &  & 1.00 & 1.00 & 1.00 &       & 1.00 & 1.00 & 0.96 \\ \bottomrule

\end{tabular}%
}
\label{table5}
\end{table}

Recall that low values of the parameter $c$ may return a large tree. In the first two scenarios, the true model does not include interaction between variables, so low $c$ parameter values return a too high Type I error. In the third scenario, the true model refers to a tree of minimum size
with a single interaction. For this reason, as the effect size of the covariates and the population size increase, higher values of parameter $c$ are required to obtain a high power. It follows that the ability of the BTRT model to find the right interactions between covariates increases when the number of judges and objects increases. In addition, if the judges’ characteristics have a high impact on the choices, then the quality of performance of the BTRT model improves considerably.

Summarizing, results of the simulation study show that a value of the pruning parameter $c$ between $0.5$ and $1$ is a good choice in almost all situations. These results are consistent with those reported in \cite{dusseldorp2010combining} for the linear regression model and in \cite{CD17} for the logistic regression model.

\section{Application on a real data set}
\label{RD}

In this section, we show a practical application of the regression trunk for preference rankings on a real data set following two different approaches. The STIMA algorithm based on the BTRT model has been implemented in the \textit{R} environment \citep{R21} by using the packages \emph{prefmod} \citep{hatzinger2012} and \emph{BradleyTerry2} \citep{turner2012}. 

The analyzed data have been collected through a survey carried out at University of Cagliari (Italy). In particular, 100 students ($H = 100$) enrolled in the first year of Master Degree in Business Economics were asked to order five characteristics of an ideal professor ($n_o = 5$) based on what they considered the most relevant: clarity of exposition ($o_1$), availability of teaching material before the lectures ($o_2$), scheduling of midterm tests ($o_3$), availability of slides and teaching material accompanying the selected books ($o_4$), helpfulness of the professor ($o_5$). 
These characteristics were ranked with values from 1 to 5, where 1 was assigned to the characteristic considered as the most important, and 5 to the least important one. Students were not allowed to indicate ties. Moreover, for each student, seven subject-specific covariates have been collected: year of study ($x_1$), total number of ECTS obtained ($x_2$), grade point average ($x_3$), course attendance in percentage ($x_4$), daily study hours ($x_5$), gender ($x_6$), and age ($x_7$). Table \ref{table6} reports the key statistics for each subject-specific covariate.

\begin{table}[ht]
\centering
\caption{Descriptive statistics of the subject-specific covariates in application.}
\resizebox{\columnwidth}{!}{
\begin{tabular}{rrrrrrrrrrrrrr}
  \hline
 & vars & n & mean & sd & median & trimmed & mad & min & max & range & skew & kurtosis & se \\ 
  \hline
Year of study &   $x_1$ & 100 & 1.18 & 0.39 & 1.00 & 1.10 & 0.00 & 1.00 & 2.00 & 1.00 & 1.64 & 0.70 & 0.04 \\ 
ECTS &  $x_2$ & 100 & 37.69 & 40.22 & 27.00 & 28.89 & 5.93 & 0.00 & 163.00 & 163.00 & 1.90 & 2.23 & 4.02 \\ 
 Grade point average &   $x_3$ & 100 & 23.02 & 6.93 & 24.80 & 24.49 & 3.26 & 0.00 & 30.00 & 30.00 & -2.36 & 5.17 & 0.69 \\ 
Course attendance &   $x_4$ & 100 & 87.37 & 13.34 & 90.00 & 89.53 & 13.34 & 40.00 & 100.00 & 60.00 & -1.22 & 0.93 & 1.33 \\ 
Daily study hours &   $x_5$ & 100 & 3.73 & 1.62 & 4.00 & 3.64 & 1.48 & 0.25 & 8.00 & 7.75 & 0.48 & 0.05 & 0.16 \\ 
Gender &   $x_6$ & 100 & 1.44 & 0.50 & 1.00 & 1.42 & 0.00 & 1.00 & 2.00 & 1.00 & 0.24 & -1.96 & 0.05 \\ 
Age &   $x_7$ & 100 & 21.00 & 3.25 & 20.00 & 20.27 & 1.48 & 19.00 & 41.00 & 22.00 & 3.16 & 13.59 & 0.33 \\ 
   \hline
\end{tabular}
\label{table6}
}
\end{table}

To apply the Bradley-Terry model, the rankings were converted in ten paired comparisons. Dealing with a small number of judges and several covariates, each judge will likely have at least one characteristic that differs from the other judges. In this framework, for each pair of comparing objects the response variable $y$ is binary and takes values of 0 and 1. Therefore, 20 observations are obtained for each judge so that the total number of rows $n$ is equal to 2,000.     

Once the design matrix is obtained, a Poisson regression model is estimated in the root node. Next, the split search as described in Section \ref{growing} is performed. In the following, we compare the results obtained for the two splitting options currently implemented for BTRT: the OSO approach and the MS approach.

\subsection{One-Split-Only (OSO) approach}\label{oso}
Based on the OSO approach, the full tree can have a maximum number of splits equal to the number of subject-specific covariates $P$. Thus, the maximum depth regression trunk has 7 splits leading to a trunk with 8 terminal nodes whose main information is summarized in Table A1 and Figure A1 in the Appendix.

Table \ref{table7} reports the node splitting information and the deviance $D$ of the final model estimated in each node (see Equation \ref{eq11}). Notice that the deviance of the main effects model is reported in the first row of Table \ref{table7} whilst the deviance of the model including a simple dichotomous variable inducing the first split of the trunk ({\it bestsplit1}) is reported in the second row. The threshold interactions are specified starting from the third row of the table, i.e. from {\it bestsplit2} onwards.

\begin{table}[h]
\centering
\caption{Pruned regression trunk: OSO approach. The table shows the node in which the split is found, the splitting covariate, and its split point together with the deviance associated with each estimated model.}
\begin{tabular}{ccrrr}
 & & & & \\
  & {Node n.} & \multicolumn{1}{c}{{$Splitting\mbox{ } covariate$}} & \multicolumn{1}{c}{{$Split\mbox{ } Point$}} & \multicolumn{1}{c}{{$Model\mbox{ } Deviance$}} \\ \hline
{} & 1 & \multicolumn{1}{c}{{main effects (no splits)}} &  & 1115 \\
{bestsplit1} & 1 & \multicolumn{1}{c}{{$x_3$ (grade point average) }} & 27.50 & 1096 \\
{bestsplit2} & 2 & {$x_7$ (age)} & 25.00 & 1080 \\
{bestsplit3} & 4 & {$x_2$ (n. of ECTS)} & 39.29 & 1064 \\
{bestsplit4} & 5 & {$x_5$ (daily study hours)} & 4.00 & 1049 \\
\hline
\end{tabular}
\label{table7}
\end{table}

The maximum-depth regression trunk is pruned applying the $c \cdot SE$ rule described in Section \ref{pruning} based on both the case-wise 10-fold cross-validation deviance ($D^{cv}$) introduced in Equation \ref{eq:state-space&obs-equ} and its standard error ($SE^{cv}$, Equation \ref{eq15}). Table \ref{table8} shows the results of the cross-validation estimates. 

\begin{table}[h!]
\centering
\caption{10-fold cross-validation results with OSO approach: $D = $ model deviance (Eq. \ref{eq11}); $D^{cv} = $ casewise cross-validation deviance (Eq. \ref{eq:state-space&obs-equ}); $SE^{cv} = $ standard error of $D^{cv}$ (Eq. \ref{eq15}).}
\begin{tabular}{crrr}
\hline
 & \multicolumn{1}{c}{{$D$}} & \multicolumn{1}{c}{{$D^{cv}$}} & \multicolumn{1}{c}{{$SE^{cv}$}} \\ \hline
{mod0} & 1115 & 0.5957 & 0.0003 \\
{mod1} & 1096 & 0.5910 & 0.0004 \\
{mod2} & 1080 & 0.5870 & 0.0005 \\
{mod3} & 1064 & 0.5858 & 0.0005 \\
{mod4} & 1049 & 0.5832 & 0.0005 \\
{mod5} & 1042 & 0.5831 & 0.0006 \\
{mod6} & 1039 & 0.5876 & 0.0007 \\
{mod7} & 1057 & 0.5906 & 0.0007 \\ \hline
\end{tabular}
\label{table8}
\end{table}

\begin{figure}[h!]
\centering
\includegraphics[width=\textwidth, height = 8cm]{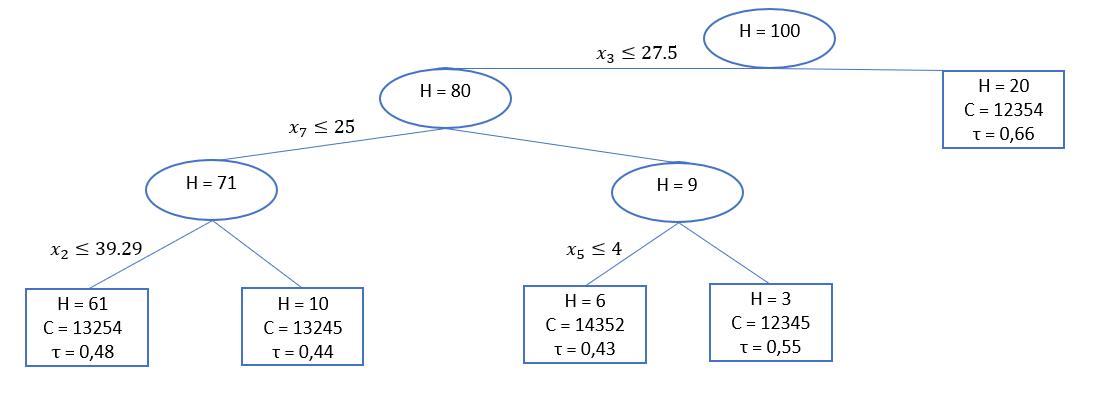}
\caption{Pruned regression trunk: OSO approach}
\label{fig2}
\end{figure}

Note that $D^{cv}$ is much smaller than the model deviance $D$, cause we used two different specifications for these two (see Equation \ref{eq11} and \ref{eq:state-space&obs-equ}): $D$ decreases between one model and another, whilst $D^{cv}$ is decreasing up to the model 5 having six terminal nodes but, from model 5 onwards, it starts to increase. Thus, using a $c \times SE$ rule with $c = 0$ leads to a pruned trunk that corresponds {\it mod5} in Table \ref{table8} and {\it bestsplit5} in Table A1 in the Appendix. Using the information obtained from the simulation study presented in Section \ref{sim}, with $n_o = 5$ and $H=100$  a possible pruning parameter is $c = 0.5$ so that the trunk is pruned starting from the fifth split ($mod5$). The final tree including four splits and $T = 5$ terminal nodes is shown in Figure \ref{fig2}.

Figure \ref{fig2}  shows the maximum-depth regression trunk. It reports the number of judges $h_t$ belonging to each node $t$. The consensus ranking $C$ is computed by using the differential evolution algorithm for median ranking detection \citep{d2017differential} and the extended correlation coefficient $\tau_x$ \citep{emond2002} within the group. Both measures are computed using the R package \emph{ConsRank} \citep{d2019consrank}. The consensus ranking reports the values associated with the objects ordered from $o_1$ to $o_5$. Ties are allowed only for the consensus ranking within the groups so that two tied objects have the same associated value. 

\subsection{Multiple Splitting (MS) approach}
The MS approach allows considering for the split search also the covariates already used in previous splits. To compare the MS approach with the OSO one, a regression trunk with the same number of terminal nodes of the OSO trunk is grown for the MS case ($T$ = 8). Results of the full tree are reported in Table A2 and Figure A2 in the Appendix. The results associated with the pruned tree are reported in Table \ref{table9}. Note that in this case the STIMA algorithm returns a trunk in which only $x_3$, $x_7$, and $x_2$ are used as splitting covariates. 

\begin{table}[h]
\centering
\caption{Pruned regression trunk: MS approach. The table shows the node in which the split is found, the splitting covariate, and its split point together with the deviance associated with each estimated model.}
\begin{tabular}{cccrr}
\hline
 & {$Node$} & {$Covariate$} & \multicolumn{1}{c}{{$Point$}} & \multicolumn{1}{c}{{$Deviance$}} \\ \hline
{} & 1 & \multicolumn{1}{c}{{main effects (no splits)}} &  & 1115 \\ 
{bestsplit1} & root & $x_3$ (grade point average) & 27.50 & 1096 \\
{bestsplit2} & 2 & $x_7$ (age) & 25.00 & 1080 \\
{bestsplit3} & 5 & $x_3$ (grade point average) & 22.00 & 1057 \\
{bestsplit4} & 4 & $x_2$ (n. of ECTS) & 39.26 & 1036 \\
{bestsplit5} & 9 & $x_2$ (n. of ECTS) & 141.00 & 1020 \\
{bestsplit6} & 18 & $x_2$ (n. of ECTS) & 114.00 & 1007 \\ \hline
\end{tabular}
\label{table9}
\end{table}

Next, the pruning procedure is performed using once again the ten fold cross-validation estimation of the deviance and its standard error. Table \ref{table10} shows the results associated with the pruned trunk deriving from the MS approach. 

\begin{table}[h!]
\centering
\caption{10-fold cross-validation results with MS approach: $D = $ model deviance (Eq. \ref{eq11}); $D^{cv} = $ casewise cross-validation deviance (Eq. \ref{eq:state-space&obs-equ}); $SE^{cv} = $ standard error of $D^{cv}$ (Eq. \ref{eq15}).}
\begin{tabular}{crrr}
\hline
 & \multicolumn{1}{c}{{$D$}} & \multicolumn{1}{c}{{$D^{cv}$}} & \multicolumn{1}{c}{{$SE^{cv}$}} \\ \hline
{mod0} & 1115 & 0.5957 & 0.0003 \\
{mod1} & 1096 & 0.5910 & 0.0004 \\
{mod2} & 1080 & 0.5870 & 0.0005 \\
{mod3} & 1057 & 0.5776 & 0.0007 \\
{mod4} & 1036 & 0.5722 & 0.0008 \\
{mod5} & 1020 & 0.5676 & 0.0008 \\
{mod6} & 1007 & 0.5664 & 0.0009 \\
{mod7} & 996 & 0.5670 & 0.0009 \\ \hline
\end{tabular}
\label{table10}
\end{table}

\begin{figure}[h]
\centering
\includegraphics[scale = 0.75]{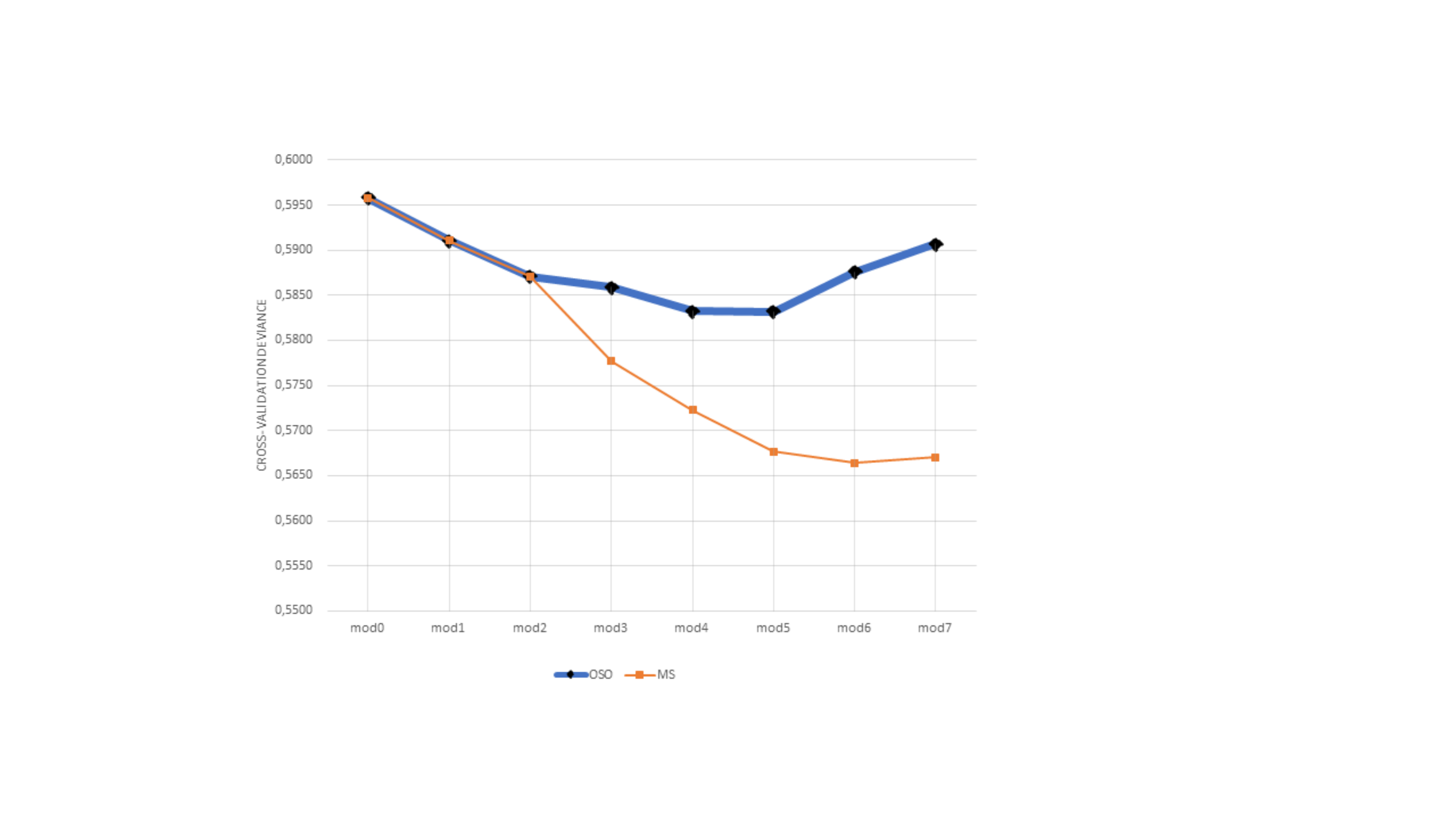}
\caption{Comparison between OSO and MS approaches}
\label{fig3}
\end{figure}

The MS approach, for each split, generates a reduction in deviance greater than that obtained with the OSO approach. The cross-validation deviance is decreasing up to model 6 ({\it mod6}), then increasing with the last split. Figure \ref{fig3} compares the two approaches in terms of cross-validation deviance obtained from one split to another. It clearly displays that the MS approach returns a regression trunk capable of better explaining the preferences expressed by the judges. 

\begin{figure}[h!]
\centering
\includegraphics[width=\textwidth, height = 8cm]{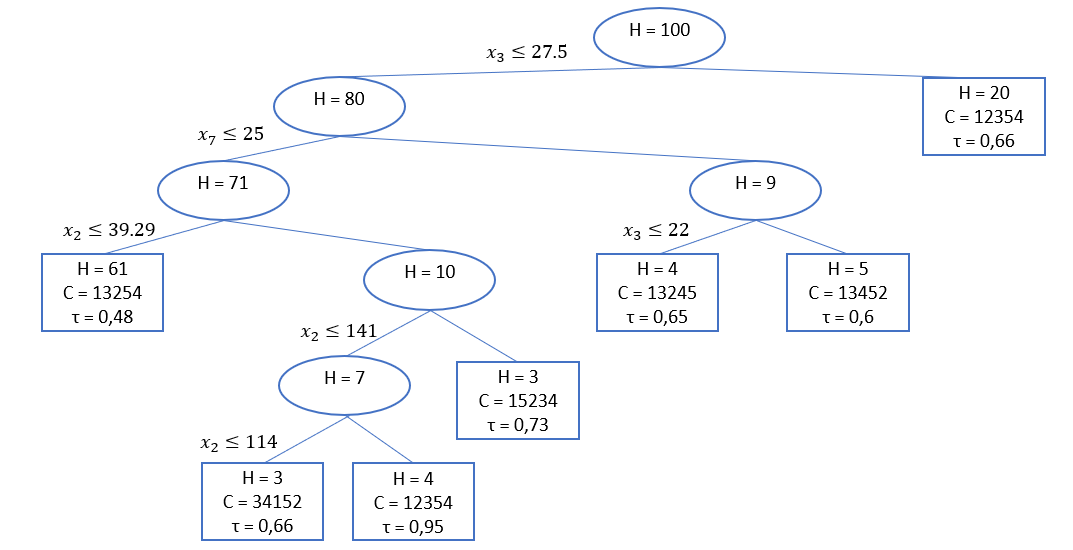}
\caption{Pruned regression trunk: MS approach}
\label{fig4}
\end{figure}

Applying the $c \cdot SE$ rule with $c = 0$ on the regression trunk grown with the MS approach  the final trunk is that corresponding to model 6 ({\it mod6}) in Table \ref{table10}. In this case, the $c \cdot SE$ rule with a value of $c$ equal to $0.5$ drives us the same pruned trunk as when $ c = 0$. Figure \ref{fig4} shows the pruned regression trunk with six splits and $T = 7$.

Note that in the pruned tree the professor's quality of exposition ($o_1$) is always preferred to all the other objects, except by the judges in Region 2. This difference in terms of consensus ranking does not emerge from the interpretation of the pruned tree obtained with the OSO approach in Figure \ref{fig2}. Region 2 is made up of students under the age of 25, with a number of ECTS less than 114 and with an average grade of less than 27.5 points.

As expected, the two approaches provide different results: the OSO approach detects the interaction between all the variables under study, but does not return the best regression trunk in terms of goodness of fit. The MS approach returns a trunk that fits the data better but the final BTRT model may be more challenging to interpret. 

The model deriving from the MS regression trunk returns the following coefficients (with standard deviations in parenthesis) estimated after setting the fifth object $o_5$ (the professor helpfulness) as the reference level, so that the estimated parameters associated to $\hat{\lambda}_{o_5,h}$ are automatically set to zero:

\begin{align*}
\hat{\lambda}_{o_1,h} & = &-\underset{(1.78)}{0.30}
					   -\underset{(0.45)}{1.24}^{***}x_1
					   +\underset{(0.01)}{0.03}^{***}x_2
					   -\underset{(0.03)}{0.04}x_3
					   -\underset{(0.01)}{0.01}x_4
					   -\underset{(0.07)}{0.01}x_5
					   +\underset{(0.19)}{0.43}^{**}x_6
					   +\underset{(0.06)}{0.18}^{***}x_7+\\  
   				     &  &- \underset{(0.29)}{0.40}R_2
				            -\underset{(0.63)}{2.99}^{***}R_3
				            +\underset{(36.87)}{3.84}R_4
				            -\underset{(0.91)}{3.85}^{***}R_5
				            -\underset{(0.88)}{1.54}^*R_6
				            -\underset{(0.79)}{4.00}^{***}R_7\\
\hat{\lambda}_{o_2,h} & = &\underset{(1.53)}{2.32}
					   -\underset{(0.44)}{0.79}x_1
					   +\underset{(0.01)}{0.02}^{***}x_2
					   -\underset{(0.03)}{0.04}x_3
					   -\underset{(0.01)}{0.02}^{**}x_4
					   -\underset{(0.06)}{0.05}x_5
					   +\underset{(0.16)}{0.36}^{**}x_6
					   +\underset{(0.05)}{0.02}x_7+\\  
   				     &  &- \underset{(0.24)}{0.20}R_2
				            -\underset{(0.99)}{0.53}R_3
				            -\underset{(0.74)}{2.20}^{***}R_4
				            +\underset{(0.78)}{0.52}R_5
				            -\underset{(0.88)}{1.54}^*R_6
				            -\underset{(0.65)}{1.87}^{***}R_7\\
\hat{\lambda}_{o_3,h} & = &\underset{(1.54)}{1.65}
					   -\underset{(0.44)}{0.50}x_1
					   +\underset{(0.01)}{0.01}^{**}x_2
					   -\underset{(0.03)}{0.00}x_3
					   -\underset{(0.01)}{0.02}^{**}x_4
					   -\underset{(0.06)}{0.12}^{**}x_5
					   +\underset{(0.16)}{0.37}^{**}x_6
					   +\underset{(0.05)}{0.01}x_7+\\  
   				     &  &- \underset{(0.24)}{0.01}R_2
				            -\underset{(0.51)}{0.74}R_3
				            +\underset{(0.88)}{0.49}R_4
				            -\underset{(0.83)}{2.20}^{***}R_5
				            +\underset{(0.76)}{0.95}R_6
				            -\underset{(0.64)}{1.83}^{***}R_7\\
\hat{\lambda}_{o_4,h} & = &-\underset{(1.71)}{3.89}^*
					   -\underset{(0.46)}{0.91}^{**}x_1
					   +\underset{(0.01)}{0.02}^{***}x_2
					   +\underset{(0.03)}{0.06}^*x_3
					   -\underset{(0.01)}{0.01}^{*}x_4
					   -\underset{(0.06)}{0.09}x_5
					   +\underset{(0.17)}{0.38}^{**}x_6
					   +\underset{(0.06)}{0.14}^{**}x_7+\\  
   				     &  &+ \underset{(0.26)}{0.67}^{***}R_2
				            -\underset{(0.53)}{0.55}R_3
				            +\underset{(0.80)}{0.24}R_4
				            -\underset{(0.73)}{0.65}R_5
				            +\underset{(0.77)}{0.92}R_6
				            -\underset{(0.67)}{2.15}^{***}R_7,
\end{align*}

The stars $'*'$ associated to some estimated coefficients indicate that they are significantly different from zero with a pvalue lower than $0.001$ ($'***'$), $0.01$ ($'**'$) and $0.05$ ($'*'$), respectively.
The fifth object $o_5$ (the professor helpfulness) is treated as reference level, so that the estimated parameters are automatically set to 0.  The regions $R_2,\ldots,R_7$ obtained from the regression trunk represented in Figure \ref{fig4} are defined as follows:

\begin{align*}
&R_2 = I(x_3\leq27.5, x_7\leq25, x_2\leq39.29),\\
&R_3 = I(x_3 \leq 27.5, x_7 \leq 25, x_2\leq 114),\\
&R_4 = I(x_3 \leq 27.5, x_7 \leq 25, x_2\geq 114),\\
&R_5 = I(x_3 \leq 22.5, x_7 > 25, x_2 >141),\\
&R_6 = I(x_3 \leq 22, x_7 >25),\\
&R_7 = I( 22 < x_3 \leq 27.5).\\
\end{align*}

The region $R_1$ plays the role of reference category. It is defined by the indicator function $I(X_3 > 27.5)$. From the main effects side, the final model shows that the covariates $x_2$ (total number of ECTS achieved) and $x_6$ (gender) have a significant and positive effect on the preferences expressed about each object. In particular, looking at the $\beta_{i,6}$ coefficients, it can be seen that as the number of ECTS obtained increases, the tendency to prefer the professor's clarity ($o_1$) to his helpfulness ($o_5$) is slightly higher. On the contrary, looking at the effect of the year of enrollment ($x_1$) on the preference for the professor's clarity,  it seems that the higher the enrollment year, the lower the tendency to prefer this attribute to the professor's helpfulness. 
These two results seem to be in contrast with each other, but in reality they highlight the fact that the year of enrollment and the number of ECTS acquired are two covariates that provide different information about students.

As for the interaction effects, the last region $R_7$ shows significant and negative coefficients whatever the considered object. In each case, when the students' grade point average is between 22 and 27.5, there is a strong tendency to prefer the professor helpfulness to all other attributes.

\section{Conclusions}
\label{conc}
This paper introduces a new Bradley-Terry Regression Trunk (BTRT) model to analyze preference data. BTRT is based on a probabilistic approach in which the judges’ heterogeneity is taken into account with the introduction of subject-specific covariates. 

The combination of the log-linear Bradley-Terry model with the regression trunk methodology allows to generate, through Poisson regressions, an easy to read partition of judges based on their characteristics and the preferences they have expressed. 

The main effects on the object choice of the judges’ characteristics and their interactions are simultaneously estimated. BTRT accounts for the drawback of to the classic tree-based models when no a priori hypotheses on the interaction effects are available. At the same time, it allows to detect threshold interactions in an automatic and data-driven mode. The final result is a small and easily interpretable tree structure, called regression trunk, that only considers the interactions that bring significant improvements to the main effects model fit.

Simulations showed that the ability of the BTRT mode to find the right interactions increases when both the sample size and the number of objects to be judged increase, particularly if the covariates have a high impact on the choices. The results suggest that in most of the cases a value of the pruning parameter $c$ between 0.7 and 0.9 is a good choice. These values are consistent with those reported in \cite{dusseldorp2010combining} for the linear regression model and in \cite{CD17} for the logistic regression model.

The two different approaches that have been introduced for the BTRT model have both been used in a real dataset application. It emerges that the One-Split-Only approach aims to verify the interaction effect between all the covariates taken into consideration and the final result is easier to interpret. On the other hand, the Multiple Splitting approach yields a tree more capable of capturing the most significant interactions between the variables selected by the model.

The BTRT model appears well-suited to analyze the probability distribution of preferring a particular object for a specific group of individuals with a specific set of characteristics. For this reason, it can be used for both descriptive and predictive purposes as it allows the user to estimate the impact of each subject-specific covariate on the judges' choices, the overall consensus ranking, and the effect size of the interactions between covariates. 

Future research is addressed to consider cases when categorical subject-specific covariates with more than two categories are used as possible split candidates as well as to investigate further model performance and stability with respect to (big) datasets presenting a high number of objects, rankings, and covariates. This would allow to better evaluate the two approaches illustrated in Section \ref{RD}. 

At the same time, research efforts will be aimed at extending the model to cases where ties (i.e., weak orderings) or missing values (i.e., partial orderings) are allowed. Future research may also be oriented to the extension of the BTRT model for the analysis of ordinal data treated as rankings, using not only information relating to the judges, but also the characteristics of the objects themselves (i.e., object-specific covariates).

\vspace{\fill}\clearpage

\bibliographystyle{newapa}

\newpage

\section*{Appendix}

\begin{table}[h!]
\centering
\caption{A1. Full regression trunk: OSO approach. The table shows the node in which the split is found, the splitting covariate, and its split point together with the deviance associated with each estimated model.}
\begin{tabular}{ccrrr}
 & & & & \\
  & {Node n.} & \multicolumn{1}{c}{{$Splitting\mbox{ } covariate$}} & \multicolumn{1}{c}{{$Split\mbox{ } Point$}} & \multicolumn{1}{c}{{$Model\mbox{ } Deviance$}} \\ \hline
{} & 1 & \multicolumn{1}{c}{{main effects (no splits)}} &  & 1115 \\
{bestsplit1} & 1 & \multicolumn{1}{c}{{$x_3$ (grade point average) }} & 27.50 & 1096 \\
{bestsplit2} & 2 & {$x_7$ (age)} & 25.00 & 1080 \\
{bestsplit3} & 4 & {$x_2$ (n. of ECTS)} & 39.29 & 1064 \\
{bestsplit4} & 5 & {$x_5$ (daily study hours)} & 4.00 & 1049 \\
{bestsplit5} & 3 & {$x_4$ ($\%$ course attendance)} & 99.00 & 1042 \\
{bestsplit6} & 9 & {$x_6$ (gender)} & 1.00 & 1037 \\ 
{bestsplit7} & 10 & {$x_1$ (year of study)} & 1.00 & 1039 \\
\hline
\end{tabular}
\label{table11}
\end{table}

\begin{figure}[h!]
\centering
\includegraphics[width=\textwidth, height = 8cm]{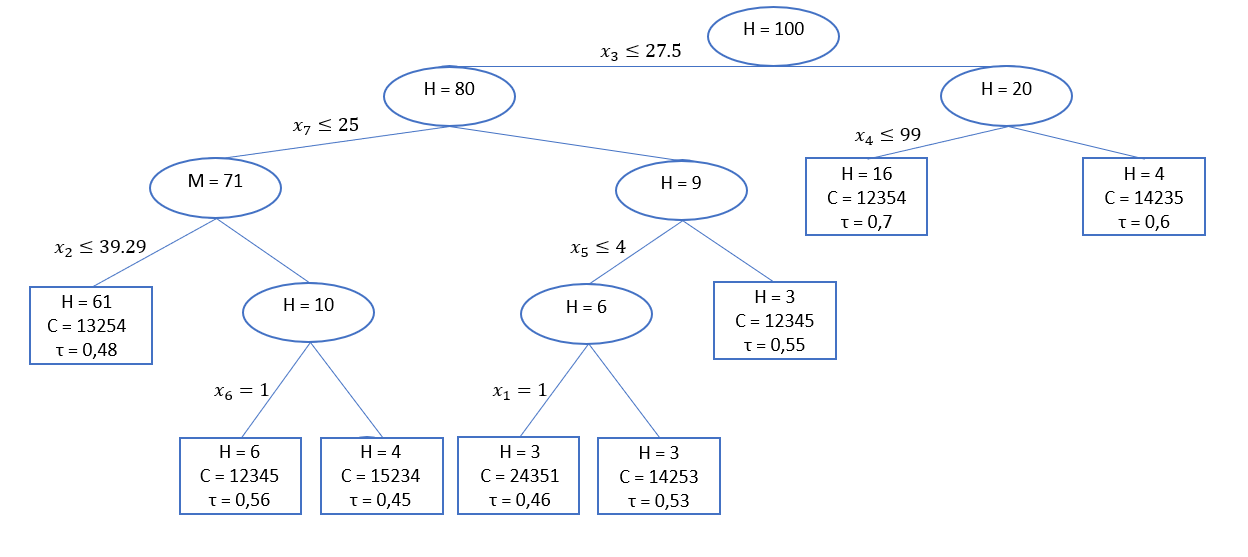}
\caption{A1. Full regression trunk: OSO approach} 
\label{fig5}
\end{figure}

\begin{table}[h]
\centering
\caption{A2. Full regression trunk: MS approach. The table shows the node in which the split is found, the splitting covariate, and its split point together with the deviance associated with each estimated model.}
\begin{tabular}{cccrr}
\hline
 & {$Node$} & {$Covariate$} & \multicolumn{1}{c}{{$Point$}} & \multicolumn{1}{c}{{$Deviance$}} \\ \hline
{} & 1 & \multicolumn{1}{c}{{main effects (no splits)}} &  & 1115 \\ 
{bestsplit1} & root & $x_3$ (grade point average) & 27.50 & 1096 \\
{bestsplit2} & 2 & $x_7$ (age) & 25.00 & 1080 \\
{bestsplit3} & 5 & $x_3$ (grade point average) & 22.00 & 1057 \\
{bestsplit4} & 4 & $x_2$ (n. of ECTS) & 39.26 & 1036 \\
{bestsplit5} & 9 & $x_2$ (n. of ECTS) & 141.00 & 1020 \\
{bestsplit6} & 18 & $x_2$ (n. of ECTS) & 114.00 & 1007 \\
{bestsplit7} & 8 &  $x_3$ (grade point average) & 24.49 & 996 \\ \hline
\end{tabular}
\label{table12}
\end{table}

\begin{figure}[h]
\centering
\includegraphics[width=\textwidth, height = 8cm]{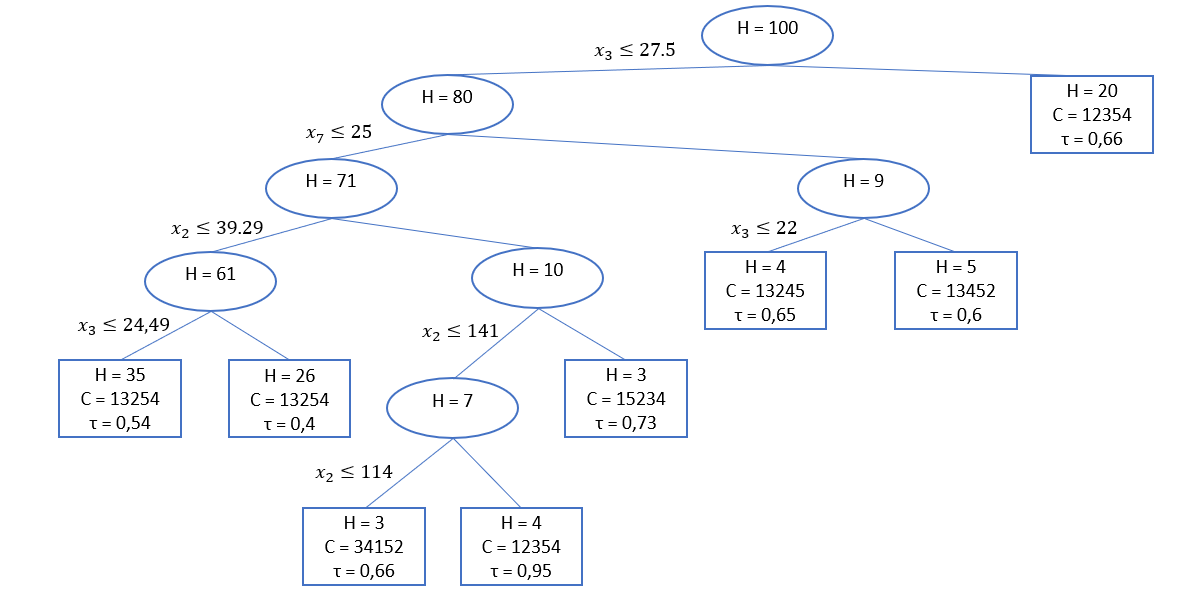}
\caption{A2. Full regression trunk: MS approach}
\label{fig6}
\end{figure}

\end{document}